# Optimizing the Use of an Artificial Tongue-Placed Tactile Biofeedback for Improving Ankle Joint Position Sense in Humans

N. Vuillerme, O. Chenu, A. Fleury, J. Demongeot and Y. Payan
Laboratoire TIMC-IMAG, UMR CNRS 5525, La Tronche, France

*Abstract* — The performance of an artificial tongue-placed tactile biofeedback device for improving ankle joint position sense was assessed in 12 young healthy adults using an active matching task. The underlying principle of this system consisted of supplying individuals with supplementary information about the position of the matching ankle relative to the reference ankle position through a tongue-placed tactile output device generating electrotactile stimulation on a 36-point (6 × 6) matrix held against the surface of the tongue dorsum. Precisely, (1) no electrodes were activated when both ankles were in a similar angular position within a predetermined "angular dead zone" (ADZ); (2) 12 electrodes (2 × 6) of the anterior and posterior zones of the matrix were activated (corresponding to the stimulation of the front and rear portion of the tongue) when the matching ankle was in a too plantarflexed and dorsiflexed position relative to the reference ankle, respectively. The effects of two ADZ values of 0.5° and 1.5° were evaluated. Results showed (1) more accurate and more consistent matching performances with than without biofeedback and (2) more accurate and more consistent ankle joint matching performances when using the biofeedback device with the smaller ADZ value. These findings suggest that (1) electrotactile stimulation of the tongue can be used to improve ankle joint proprioception and (2) this improvement can be increased through an appropriate specification of the ADZ parameter provided by the biofeedback system. Further investigations are needed to strengthen the potential clinical value of this device.

## I. INTRODUCTION

IN the last decades, a growing number of tactile displays have been developed for sensory substitution or augmentation systems (e.g., [1-4]). These devices are designed to evoke tactile sensation within the skin at the location of the tactile stimulator using either mechanical ("vibrotactile display") or electrical ("electrotactile display") stimulation. The skin being the interface between the tactile stimulator and the neural sensory system, the performance of a tactile display largely depends on the neurophysiologic characteristics of the receptive body regions. Along these lines, the human tongue has recently been suggested to provide a promising electrotactile stimulation site [5]. Because of its dense mechanoreceptive innervations [6] and large somatosensory cortical representation [7], the tongue is indeed a very sensitive organ which can convey higher-resolution information than the skin can [8-9]. In addition, due to the excellent conductivity offered by the saliva, electrotactile stimulation of the tongue requires only 3% of the voltage (5 - 15 V) and much less current (0.4 - 2.0 mA) than those required for the fingertip [5]. The Tongue Display Unit (TDU) has recently been developed to take advantage of these characteristics [5]. Initially used as a tactile-vision sensory substitution system to provide distal spatial information to blind people (e.g., [1,8,10,11]), the TDU also has proven its efficiency as a tactile-proprioception sensory augmentation system [12]. The performance of this system ("TDU-biofeedback system") in improving proprioceptive acuity at the ankle joint recently has been evaluated on young healthy adults using an active matching task.

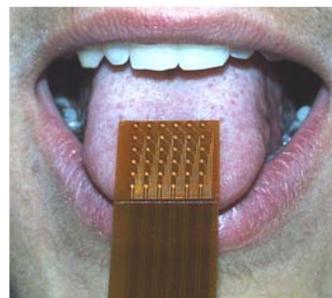

Fig. 1. Photograph of the Tongue Display Unit used in the present experiment. It comprises a 2D electrode array (1.5 × 1.5 cm) consisting of 36 gold-plated contacts each with a 1.4 mm diameter, arranged in a 6 × 6 matrix.

Position of the matching ankle relative to the reference ankle was fed back to a tongue-placed tactile output device generating electrotactile stimulation on a 36-point (6 × 6) matrix held against the surface of the tongue dorsum (Fig. 1) as follows: (1) no electrodes were activated when both ankles were in a similar angular position within a predetermined "angular dead zone" (ADZ); and (2) 12 electrodes (2 × 6) of the anterior or posterior zones of the matrix were activated (corresponding to the stimulation of the front and rear portion of the tongue) when the matching ankle was in a too plantarflexed and dorsiflexed position relative to the reference ankle, respectively. Results of this experiment showed more accurate and more consistent matching performances when using the TDU-biofeedback system with an ADZ equal to 1,5°. At this point, it is

Nicolas Vuillerme is with the TIMC-IMAG Laboratory, Equipe AFIRM, UMR CNRS 5525, Faculté de Médecine de Grenoble, Bâtiment Jean Roget, F38706 La Tronche Cédex; phone: +33 4 76 63 74 86; fax: +33 4 76 51 86 67; mail: Nicolas.Vuillerme@imag.fr

Olivier Chenu, Anthony Fleury, Jacques Demongeot and Yohan Payan are also with the TIMC-IMAG Laboratory, UMR CNRS 5525, La Tronche, France. Mail: firstname.lastname@imag.fr

possible that the observed effects markedly depend on the specification of the ADZ value. The purpose of the present study was thus designed to assess this possibility, by comparing the effects of providing two different ADZ values (0.5° and 1.5°) on the performance of the TDU-biofeedback system. It was hypothesized that (1) biofeedback delivered through the TDU would improve position sense at the ankle joint and (2) that this improvement would be larger with smaller ADZ value (0.5°). Note that such a result could provide further insight into the conception and optimization of rehabilitative programs or ergonomical assistive tools.

## II. METHODS

### A. Participants

Twelve young healthy males adults (age = 27.7 ± 5.4 years; body weight = 68.9 ± 6.9 kg; height = 177.7 ± 6.5 cm) volunteered for this study and gave their informed consent to the experimental procedure as required by the Helsinki declaration (1964) and the local Ethics Committee. None of the participants presented any history of injury, surgery or pathology to either lower extremity that could affect their ability to perform the experiment.

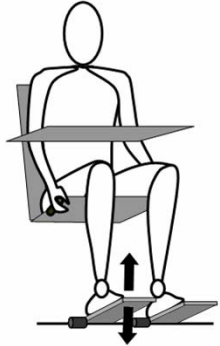

Fig. 2. Illustration of the experimental setup used for measuring ankle joint position sense.

### B. Apparatus for measuring ankle joint position sense

Participants were seated barefoot in a chair with their right and left foot secured to two rotating footplates. The knee joints were flexed at about 110°. Movement was restricted to the ankle in the sagittal plane, with no movement occurring at the hip or knee. The axes of rotation of the footplates were aligned with the axes of rotation of the ankles. Precision linear potentiometers attached on both footplates provided analog voltage signals proportional to the ankles' angles. A handheld press-button allowed recording the matching. Signals from the potentiometers and the press-button were sampled at 100 Hz (12 bit A/D conversion), then processed and stored within the Labview 5.1 data acquisition system. In addition, a panel was placed above the subject's legs to eliminate visual feedback about both ankles position (Fig. 2).

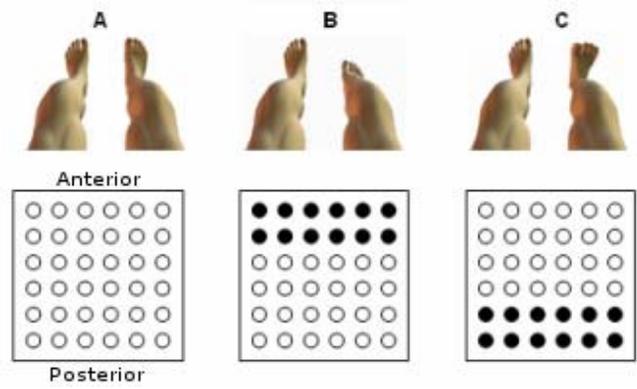

Fig. 3. Sensory coding schemes for the TDU (lower panels) as a function of the position of the matching right ankle relative to the reference left ankle (upper panels). Black dots represent activated electrodes. A: no electrodes are activated when both ankles are in a similar angular position within a range of 0.5° and 1.5°, for the $TDU_{0.5°}$ and $TDU_{1.5°}$ conditions, respectively. B: 12 electrodes (2 × 6) of the anterior zone of the matrix are activated (corresponding to the stimulation of the front portion of the tongue dorsum) when the matching right ankle was in a too plantarflexed position relative to the reference left ankle. C: 12 electrodes (2 × 6) of the posterior zone of the matrix are activated (corresponding to the stimulation of the rear portion of the tongue dorsum) when the matching right ankle was in a too dorsiflexed position relative to the reference left ankle.

### C. Apparatus for providing biofeedback: the TDU system

The underlying principle of the biofeedback device used in the present experiment consisted of supplying participants with supplementary information about the position of the matching right ankle relative to the reference left ankle position through a tongue-placed tactile output device [12]. Electrotactile stimuli were delivered to the front part of the tongue dorsum via flexible electrode arrays placed in the mouth, with connection to the stimulator apparatus via a flat cable passing out of the mouth. The system comprises a 2D electrode array (1.5 × 1.5 cm) consisting of 36 gold-plated contacts each with a 1.4 mm diameter, arranged in a 6 × 6 matrix [12] (Fig. 1).

The following coding scheme for the TDU was used: (1) no electrical stimulation when both ankles were in a similar angular position within ADZs of 0.5° and 1.5°, for the $TDU_{0.5°}$ and $TDU_{1.5°}$ conditions, respectively (Fig. 3A); (2) stimulation of either the anterior or posterior zone of the matrix (2 × 6 matrix) (i.e. stimulation of front or rear portions of the tongue) depending on whether the matching right ankle was in a too plantarflexed or dorsiflexed position relative to the reference left ankle, respectively (Fig. 3B and 3C, respectively). The frequency of the stimulation was maintained constant at 50 Hz across participants, ensuring a sensation of the continuous stimulation over the tongue surface. Conversely, the sensitivity to the electrotactile stimulation varying not only between individuals but also as a function of location on the tongue, as reported in a preliminary experiment [12], the intensity of the electrical stimulating current was adjusted for each subject, and for each of the front and rear portions of the tongue

*D. Task and procedure*

The experimenter first placed the left reference ankle at a predetermined angle where the position of the foot was maintained by means of a support [13]. By doing so, participants did not exert any effort to maintain the position of the left reference ankle, preventing the contribution of effort cues coming from the reference ankle to the sense of position during the test [13]. Two matching angular target positions were presented: (1) 10° of plantarflexion ($P_{10°}$) and (2) 10° of dorsiflexion ($D_{10°}$). These positions were selected to avoid the extremes of the ankle range of motion to minimize additional sensory input from joint and cutaneous receptors (e.g., [14]).

Once the left foot had been positioned at the test angle ($P_{10°}$ vs. $D_{10°}$), subject's task was to match its position by voluntary placement of their right ankle. When they felt that they had reached the target angular position (i.e., when the right foot was presumably aligned with the left foot), they were asked to press the button held in their right hand, thereby registering the matched position.

Three experimental conditions were presented: (1) the No-TDU condition serving as a control condition, the (2) $TDU_{0.5°}$ and (3) $TDU_{1.5°}$ conditions in which participants performed the task using a TDU-biofeedback system, as described above (section IIC).

Five trials for each target angular position and each experimental condition were performed. The order of presentation of the two targets angular positions ($P_{10°}$ vs. $D_{10°}$) and the three experimental conditions (No-TDU vs. $TDU_{0.5°}$ vs. $TDU_{1.5°}$) was randomized. Participants were not given feedback about their performance and errors in the position of the right ankle were not corrected.

*E. Data analysis*

Two dependent variables were used to assess matching performances [15]. (1) The absolute error (AE), the absolute value of the difference between the position of the right matching ankle and the position of the left reference ankle, is a measure of the overall accuracy of positioning. (2) The variable error (VE), the variance around the mean constant error score, is a measure of the variability of the positioning. Decreased AE and VE scores indicate increased accuracy and consistency of the positioning, respectively [15].

*F. Statistical analysis*

Data obtained for AE and VE were submitted to separate 2 Targets angular positions ($P_{10°}$ vs. $D_{10°}$) × 3 Conditions (No-TDU vs. $TDU_{0.5°}$ vs. $TDU_{1.5°}$) analyses of variances (ANOVAs) with repeated measures of both factors. Post-hoc analyses (orthogonal planned comparisons) were performed whenever necessary. Level of significance was set at 0.05.

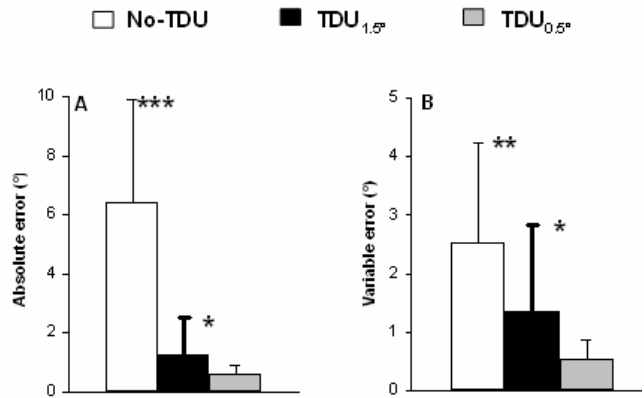

Fig. 4. Mean and standard deviation for the absolute error (A) and the variable error (B) for the three No-TDU, $TDU_{0.5°}$ and $TDU_{1.5°}$ conditions. These three experimental conditions are presented with different symbols: No-TDU (white bars), $TDU_{1.5°}$ (black bars) and $TDU_{0.5°}$ (grey bars). (*: $P < 0.05$, **: $P < 0.01$, ***: $P < 0.001$).

III. RESULTS

*A. Positioning accuracy*

Analysis of the AE showed a main effect of Condition ($F(2,22) = 39.76$, $P < 0.001$, Fig. 4A), yielding smaller values in the $TDU_{0.5°}$ and $TDU_{1.5°}$ conditions than No-TDU condition (Ps < 0.001) and smaller values in the $TDU_{0.5°}$ than $TDU_{1.5°}$ condition ($P < 0.05$). The ANOVAs showed no main effect of Target angular position ($P > 0.05$), nor any interaction of Target angular position × Condition ($P > 0.05$).

*B. Positioning variability*

Analysis of the VE showed a main effect of Condition ($F(2,22) = 16.67$, $P < 0.001$, Fig. 4B), yielding smaller values in the $TDU_{0.5°}$ and $TDU_{1.5°}$ conditions than No-TDU condition ($P < 0.001$ and $P < 0.01$, respectively) and smaller values in the $TDU_{0.5°}$ than $TDU_{1.5°}$ condition ($P < 0.05$). The ANOVAs showed no main effect of Target angular position ($P > 0.05$), nor any interaction of Target angular position × Condition ($P > 0.05$).

IV. DISCUSSION AND CONCLUSION

On the whole, more accurate and more consistent ankle joint matching performances were observed when the TDU biofeedback system was in use than when it was not. These results confirm our first hypothesis [12] and provide additional evidence that electrotactile stimulation of the tongue can be used to improve ankle joint proprioception. Interestingly, the performance of the TDU biofeedback system further was shown to depend on the specification of the ADZ parameter. Indeed, in accordance with our second hypothesis, more accurate and more consistent ankle joint matching performances were observed when using the TDU biofeedback system with the smaller ADZ value. An appropriate specification of this parameter could thus constitute a relevant means to increase the performance of the TDU biofeedback system. For instance, such a specification could allow optimizing the use of the tongue-placed tactile biofeedback device as a rehabilitation protocol or assistance based tools, by taking the patient's proprioceptive capabilities or the ongoing task's accuracy requirement into account. Note that optimizing the TDU biofeedback system is all the more relevant since accurate proprioception at the ankle joint is critical for body orientation and balance control and represents a prerequisite for different functional activities such as walking, running or driving. At this point, further investigations involving individuals with proprioceptive deficits (e.g., elderly persons, patients with diabetic neuropathy) are needed to strengthen the potential clinical value of this technique. However, these encouraging results already have led us to improve the current TDU biofeedback system by making it wireless to increase its portability to permit its use over long-time period in real-life environment. Indeed, to be acceptable as part of a viable system, this device had to be lightweight, portable, and capable of several hours of continuous operations. The current ribbon TDU system does not meet these requirements yet. Within this context, we have developed a wireless radio-controlled version of the 6 × 6 TDU matrix. This consists in a matrix glued onto the inferior part of the orthodontic retainer including microelectronics, antenna and battery, which can be worn inside the mouth like a dental retainer. This wearable device is currently tested by our industrial partner (Coronis-Systems www.coronis-systems.com) that will provide *Wavenis*, its wireless Ultra Lower Power technology which has been preferred to current radio frequency (RF) standards that have shortcomings when considering Ultra Low Power challenges.

ACKNOWLEDGMENT

We are indebted to Professor Paul Bach-y-Rita for introducing us to the TDU and for discussions about sensory substitution. The authors would like to thank subject volunteers. Special thanks also are extended to D. Flammarion and S. Maubleu for technical assistance and P. Ipou and L. Egland for various contributions.